\documentclass[pra,a4paper,superscriptaddress,twocolumn,amsmath,amssymb,floatfix,amstex]{revtex4-1}%

\usepackage{graphicx,graphics,rotating}	
\usepackage{bm}
\usepackage{hyperref}
\usepackage{color}
\usepackage{soul}

\def\hbareff{\hbar_{\mathrm{eff}}}
\def\g{\bar{g}}

\begin{document}
\title{Kapitza stabilization of a repulsive Bose-Einstein condensate \\
in an oscillating optical lattice}

\author{J.~Martin}
\affiliation{Institut de Physique Nucl\'eaire, Atomique et de
Spectroscopie, CESAM, Universit\'e de Li\`ege, B\^atiment B15, B - 4000
Li\`ege, Belgium}
\author{B.~Georgeot}
\affiliation{%
Laboratoire de Physique Th\'eorique, IRSAMC, 
Universit\'e de Toulouse, CNRS, UPS, France
}
\author{D.~Gu\'ery-Odelin}
\affiliation{%
Laboratoire Collisions, Agr\'egats, R\'eactivit\'e, IRSAMC, Universit\'e de Toulouse, CNRS, UPS, France
}
\author{D.~L.~Shepelyansky}
\affiliation{%
Laboratoire de Physique Th\'eorique, IRSAMC, 
Universit\'e de Toulouse, CNRS, UPS, France
}

\date{March 7, 2018}

\begin{abstract}
We show that the Kapitza stabilization can occur in the context of nonlinear quantum fields. Through this phenomenon, an amplitude-modulated lattice can stabilize a Bose-Einstein condensate with repulsive interactions and prevent the spreading for long times. We present a classical and quantum analysis in the framework of Gross-Pitaevskii equation, specifying the parameter region where stabilization occurs. Effects of nonlinearity lead to a significant increase of the stability domain compared with the classical case. Our proposal can be experimentally implemented with current cold atom settings.
\end{abstract}

\maketitle

\section{Introduction}

The striking example of the Kapitza pendulum shows that an oscillating force with zero average can lead to the phenomenon of Kapitza stabilization, with transformation of an unstable fixed point into a stable one \cite{kapitza1,kapitza2}. The theory of this nonlinear system is well established in a classical context \cite{landau}. 
Some applications to quantum systems have been proposed, including optical molasses \cite{surdutovich}, stability of optical resonators \cite{torosov}, trapping by laser fields \cite{ivanov,moiseyev}, cold atoms with oscillating interactions \cite{abdullaev}, the periodically driven sine-Gordon model \cite{citro}, and polariton Rabi oscillations \cite{lozovik}. 
However, the emergence of this phenomenon for nonlinear quantum fields of repulsive interactions has not been analyzed. In this paper, we show that a similar effect appears for a repulsive Bose-Einstein condensate (BEC) in an oscillating optical lattice. For this system, the oscillating lattice enables the localization of a wave packet of repulsive atoms through Kapitza stabilization: thus, while in the absence of the lattice the atoms spread over the system, they remain trapped in a localized wave packet in the presence of the oscillating force with zero mean, an effect due to the interplay between dynamical renormalization of the potential and atom-atom interactions. The evolution is described by the Gross-Pitaevskii Equation~\cite{pitaevskii} (GPE), with the repulsive nonlinear interaction creating the unstable fixed point in the vicinity of the maximum of the wave packet. In contrast with the standard classical Kapitza pendulum, where the potential is fixed in the vicinity of the unstable fixed point, the present GPE setting creates a more complex situation where the potential varies with the shape of the wave function. In the following we describe the physics of this remarkable phenomenon and present realistic parameter values for an experimental realization with a BEC. We note that the general problem of stabilization by oscillating fields finds various important applications; e.g., Paul traps for charged particles \cite{Paultrap}.

\section{Classical system dynamics}
We first analyze a classical inverted harmonic oscillator in one dimension in an oscillating periodic potential, with the Hamiltonian
\begin{equation}\label{H}
H=\frac{p^{2}}{2m} - \frac{1}{2}m\omega_{i}^{2}x^{2} + V_{0}(x)\cos(\omega_\ell t),
\end{equation}
with $V_{0}(x)=U_0\cos\left(2\pi x/d \right)$ where $U_0$ is the potential amplitude. Here $m$ is the particle mass, $x$ and $p$ are position and momentum, $\omega_{i}$ characterizes the unstable fixed point and the periodic potential has a spatial period $d$ and an amplitude oscillation of frequency $\omega_\ell$. We define a characteristic momentum $p_0=4\sqrt{mU_0}$ and oscillation frequency $\omega_0=2\pi\sqrt{U_0/(md^2)}$, leading to the dimensionless variables 
\begin{equation}
X=2\pi \frac{x}{d},\quad P=\frac{p}{p_0}, \quad T=\frac{\omega_\ell t}{2\pi}
\end{equation}
and the frequency ratios 
\begin{equation}
R_{i0}=\frac{\omega_i}{\omega_0},\quad R_{0\ell}=\frac{\omega_0}{\omega_\ell}.
\end{equation}

\begin{figure}
\includegraphics[width=1.0\linewidth]{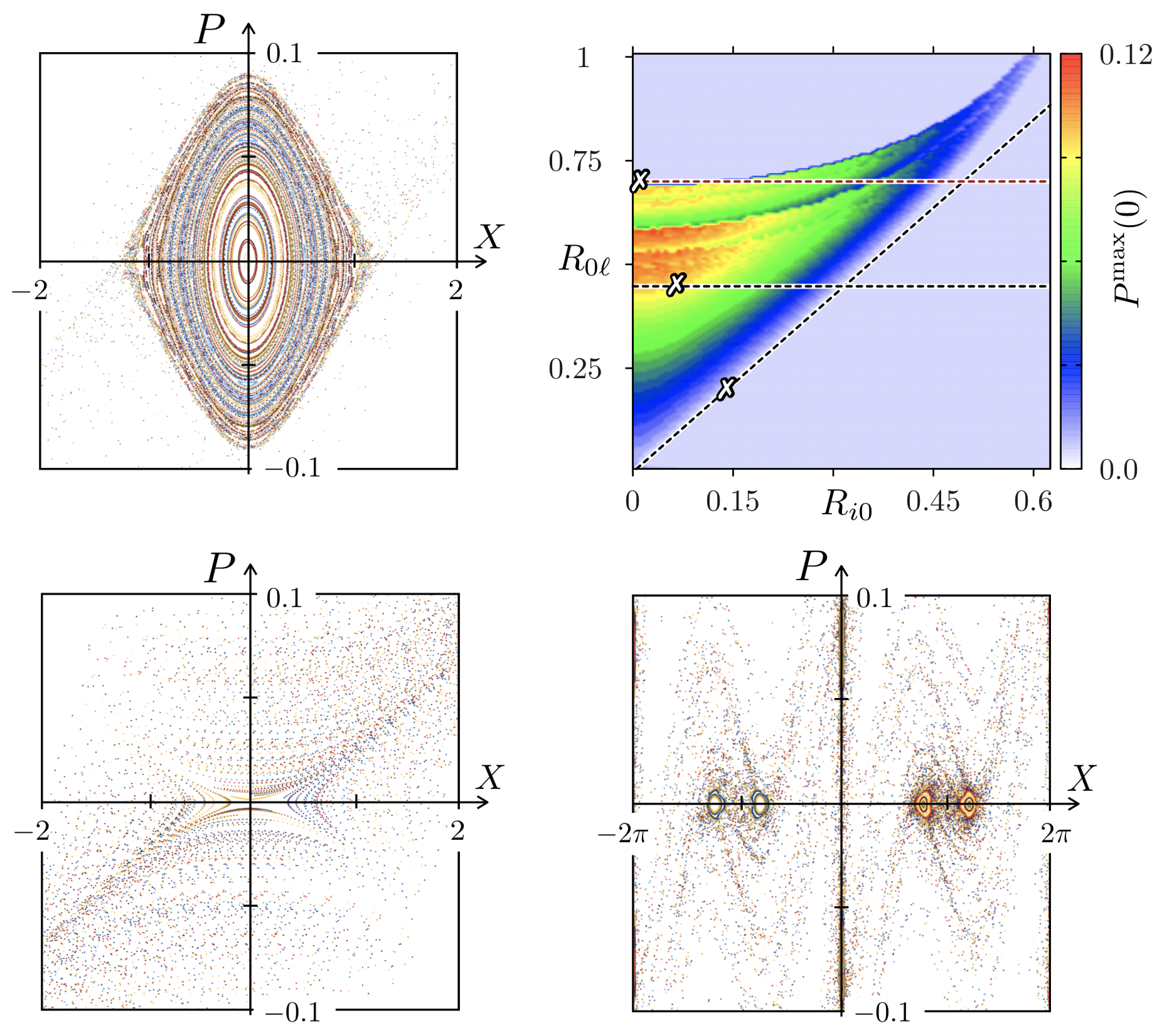}
\caption{Poincar\'e sections formed by a few thousand trajectories with random initial conditions $(X(0),P(0))\in [-2\xi:2\xi]\times [-0.1:0.1]$ ($\xi=1$ or $\pi$) propagated during a timespan $\Delta T=400$ for the frequency ratios $R_{i0}=0.075$ and $R_{0\ell}=0.45$ (top left), $R_{i0}=0.15$ and $R_{0\ell}=0.2$ (bottom left), and $R_{i0}=0.02$ and $R_{0\ell}=0.7$ (bottom right). (top right) Stability region in the parameter space of frequency ratios. Color shows the largest initial momentum $P^{\mathrm{max}}(0)$ for which trajectories with initial conditions $(X(0),P(0))=(0,P(0))$ remain stable for time $T\in[0:1000]$. Crosses mark parameters of the Poincar\'e sections; dashed black lines show theory (\ref{kapitzaborder}), upper red/grey dashed line shows refined theory taking into account secondary resonances (see text).\label{poincare_stability}}
\end{figure}

\begin{figure}
\includegraphics[width=1.0\linewidth]{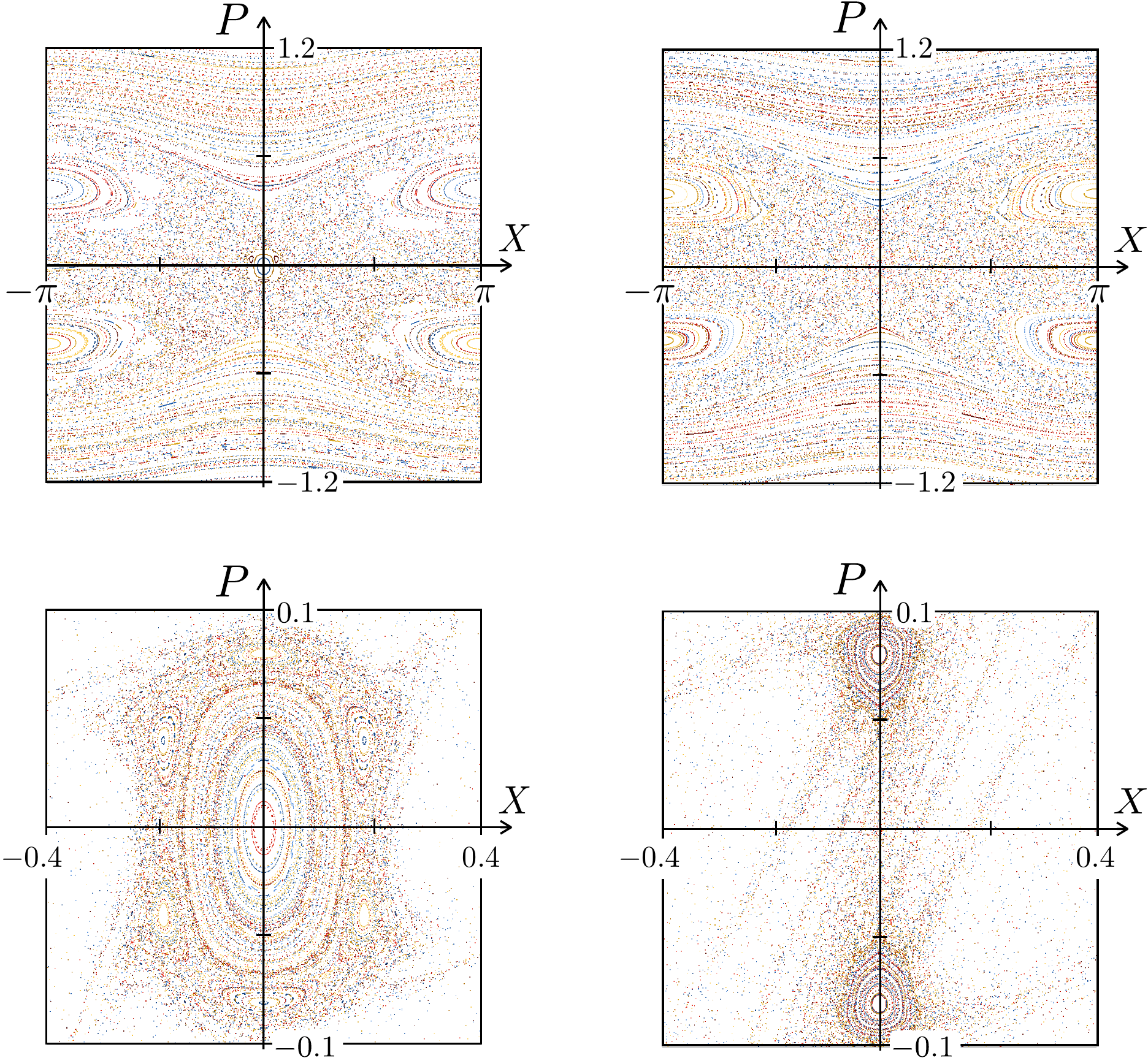}
\caption{Poincar\'e sections for $R_{0\ell}=0.65$ (left panels) and  $R_{0\ell}=0.7$ (right panels) at $R_{i0}=0$ showing the loss of stability of the fixed point at $P=X=0$.\label{poincare_stability_2}}
\end{figure}

Following the standard methods of dynamical systems \cite{lichtenberg}, we describe the dynamics through the Poincar\'e section, with typical phase-space structures shown in   
Fig.~\ref{poincare_stability}. The bottom-left panel shows the regime where the Kapitza stabilization is too weak and the point $X=P=0$ remains unstable. The top left panel shows the regime of Kapitza stabilization with a stability island around $X=P=0$; the island is surrounded by a chaotic component where the trajectories can escape to infinity. The bottom right panel corresponds to a very weak value of $R_{i0}$ and relatively strong driving, with overlapping resonances leading to onset of chaos as determined by the Chirikov criterion \cite{chirikov1979}. To determine numerically the stability diagram, we follow trajectories with random initial conditions for sufficiently long time $\Delta T$. A trajectory of initial conditions $(X(0),P(0))=(0,P(0))$ is considered unstable if $|X(T)-0|>\pi$ for some $T\in[0:1000]$. The top-right panel shows a density plot of the largest initial momentum $P(0)$ giving rise to a stable trajectory as a function of $R_{i0}$ and $R_{0\ell}$. It highlights the parameter region where the Kapitza phenomenon stabilizes the unstable fixed point. The specific shape of this region depends on two main borders, the lower one determined through Kapitza's original argument \cite{kapitza1,kapitza2,landau}, and the upper one through the Chirikov criterion \cite{chirikov1979}, yielding
\begin{equation}\label{kapitzaborder} 
R_{0\ell} > \sqrt{2} R_{i0} ,\;\; R_{0\ell} < 0.45
\end{equation}
The derivation of the first relation directly follows the approach of Kapitza pendulum \cite{landau}: the effective average potential created by the oscillating force is
$U_{\mathrm{eff}}= \langle{\dot p}^2\rangle /(2 m \omega_{\ell}^2) = 
( \pi^2 U_0^2/(m d^2 \omega_{\ell}^2)) \sin^2 (2\pi x/d)$, which combined with the inverted harmonic potential, gives for small oscillations the  squared effective frequency $\omega_{\mathrm{eff}}^2 = (\omega_0^2/\omega_{\ell})^2/2 - \omega_i^2$. Thus, $x=0$ is stable if $\omega_{\mathrm{eff}}^2>0$, leading to the first inequality of Eq.~(\ref{kapitzaborder}). The second inequality
follows from the Chirikov criterion \cite{chirikov1979}.
The nonlinear resonances are located at positions
$p_\pm =\pm \omega_{\ell} d/2\pi$. The resonant term of the
Hamiltonian
is reduced to a pendulum Hamiltonian
$H_{rs}=p^2/2m + U_0/2 \cos \theta$
with the phase $\theta = 2\pi x/d \pm \omega_\ell t$
conjugate to $p$.
According to the standard results  for a pendulum~\cite{chirikov1979},
the frequency width of the pendulum
separatrix is $\Delta \omega = 2\,\omega_0$,
and the frequency distance between resonances
is $\delta \omega = 2\, \omega_\ell$.
The parameter of Chirikov resonance overlap criterion
is $S =\Delta \omega/ \delta \omega$ and
the chaotic transitions between two resonances take place
at $K \approx 2.5\, S^2 > 1$, leading to
Eq.~({\ref{kapitzaborder}) (the coefficient 2.5 takes into account the effect of secondary resonances)
\cite{chirikov1979} (see also \cite{dlsscholar}).

These two theoretical borders of Eq.~({\ref{kapitzaborder}) are shown by straight black dashed lines in Fig.~\ref{poincare_stability}. We note that the lower border is in excellent agreement
with the numerical data. The upper border is
lower than the stability region centred around
$R_{0\ell} \approx 0.5$. The reason is that the Chirikov overlap
criterion gives the border for a chaotic transition between resonances
while the destruction of
resonances takes place at higher values
(e.g.\ in the Chirikov standard map,
the primary resonance becomes unstable at $K \approx 2.5 S^2 =4$
while the last invariant curve is destroyed at
$K \approx 0.9716$). In our case with two primary
resonances, the secondary resonance at $P=X=0$
becomes unstable at $R_{0\ell} \approx 0.7$ and $R_{i0}=0$,
as it is shown in the Poincare sections in Fig.~\ref{poincare_stability_2}
for $R_{0\ell}=0.65$ (fixed point $P=X=0$ is stable)
and  $R_{0\ell}=0.7$ (fixed point $P=X=0$ is unstable)
at  $R_{i0}=0$. The refined upper stability border $R_{0\ell}=0.7$ is shown in top right panel of Fig.~\ref{poincare_stability} by the upper horizontal red dashed line.

\section{Quantum evolution with GPE}
We now turn to the quantum case, and set $U_0=s\,E_{L}/2$ where $s$ is a dimensionless parameter characterizing the lattice depth and $E_{L}=2\pi^{2}\hbar^{2}/(md^{2})$ is a lattice characteristic energy~\cite{For16}. When the dimensionless position and momentum are turned into operators, $\hat{X}=X$ and $\hat{P}=(\hbareff/i)\partial_X$, the canonical commutation relation $[\hat{x},\hat{p}]=i\hbar$ leads to an effective Planck's constant $\hbareff=1/(2\sqrt{s})$.

In this work, we use the one-dimensional (1D) GPE to study the dynamics of a BEC with $N$ atoms subjected to the driving potential $V_{0}(x)\cos(\omega_\ell t)$. This equation is valid in the weakly interacting regime. More quantitatively, this regime is obtained from an anisotropic three-dimensional (3D) confinement tightly confined in the radial direction, i.e., when the confinement energy, $\hbar \omega_\perp$, greatly exceeds the mean-field interaction energy. As a result, at sufficiently low temperature the radial motion is frozen and therefore governed by the ground state wave function of the radial harmonic oscillator. The strength of the interaction in the GPE equation is then renormalized by averaging the 3D interaction over the radial density profile: $g=2\hbar^2a_s/m\ell_\perp^2$, where $a_s$ is the 3D scattering length and $\ell_\perp=(\hbar/m\omega_\perp)^{1/2}$ the harmonic oscillator length associated with the radial harmonic confinement. In 1D, the criterion of the weakly interacting regime reads $\gamma = mg/\hbar^2 n \ll 1$ where $n$ is the 1D atomic density \cite{olshani98,petrov04}. In the weakly interacting regime, the BEC wave function (normalized according to $\int|\psi|^{2}dx=\int|\Psi|^{2}dX=1$ where $\Psi\equiv \sqrt{d/2\pi}\,\psi$) is thus governed by the nonlinear equation
\begin{equation}\label{GPE}
\frac{i}{4\pi}\partial_T \Psi=R_{0\ell}\left(-\hbareff\partial_X^2+\frac{\cos(2\pi T)\cos X}{8\hbareff}+\frac{\g |\Psi|^{2}}{2}\right)\Psi
\end{equation}
where
$\g=2\pi Ng_{1D}/(\hbar\omega_0 d)=4\pi N \omega_{\perp}a_{s}/(\omega_{0} d)$ \cite{Jac98}.

By expanding the nonlinear potential for a Gaussian wave packet $|\psi(x)|^2=\exp [-x^2/(2\sigma^2)]/(\sigma \sqrt{2\pi})$ of rms width $\sigma$ around its maximum, we obtain an effective inverted harmonic potential with rescaled frequency
\begin{equation}\label{effRi0}
R_{i0,\mathrm{eff}}\equiv \frac{\omega_{i,\mathrm{eff}}}{\omega_0}= 2^{3/4} \pi^{-1/4} \sqrt{\frac{\hbar_{\mathrm{eff}}\g}{\tilde{\sigma}^3}}
\end{equation}
where $\tilde{\sigma}=2\pi\sigma/d$. From the expression for $U_{\mathrm{eff}}$, it follows that 
all points at $X = m\pi$ with integer $m$ become stable for $R_{i0,\mathrm{eff}} < R_{0\ell}/\sqrt{2}$.

In our simulations, we use the Strang-Marchuk operator-splitting method \cite{strang} to approximate the evolution operator corresponding to Eq.~(\ref{GPE}). We take $N_\mathrm{s}=2^{16}$ basis states for the wave function, a range of $X$ values corresponding to $Q_{\mathrm{tot}}=64$ periods of the driven optical lattice [which leads to a numerical grid with $\delta X=(2\pi Q_{\mathrm{tot}})/N_\mathrm{s}\approx 0.006$ and $\delta P=\hbar_{\mathrm{eff}}/Q_{\mathrm{tot}}\approx 5.5\times 10^{-4}$ for $s=200$], and a time step $\delta T\in [0.0002:0.001]$. If $Q$ denotes the total number of initially populated potential wells of the static potential $-V_0(X)$, then the effects of interactions in our simulations only depend on the ratio $\g/Q$ since the nonlinear potential is given by $\g|\Psi|^2$. In the case of one localized packet ($Q=1$), the initial state is the ground state (without atom-atom interactions) of the potential well centered at $X=0$ of the static potential $-V_0(X)$, which for large enough $s$ corresponds to a Gaussian wave packet $|\Psi(X)|^2=\exp [-X^2/(2\tilde{\sigma}^2)]/(\tilde{\sigma} \sqrt{2\pi})$ of rms width $\tilde{\sigma}=1/s^{1/4}$ and zero average momentum. For such initial states, we have $ R_{i0,\mathrm{eff}}=(2/\pi)^{1/4} s^{1/8} \sqrt{\g}$. We also consider as initial state a chain of wave packets periodically repeated in potential wells ($Q=Q_{\mathrm{tot}}$), which we describe as the ground state (with atom-atom interactions) of GPE for the static potential $-V_0(X)$. Such states can be easily prepared experimentally by switching a static optical lattice with a formation of BEC in each potential minimum. For such an initial state one should replace $\g$ by $\g/Q_{\mathrm{tot}}$ in Eq.~(\ref{effRi0}).

\begin{figure}
\includegraphics[width=\linewidth]{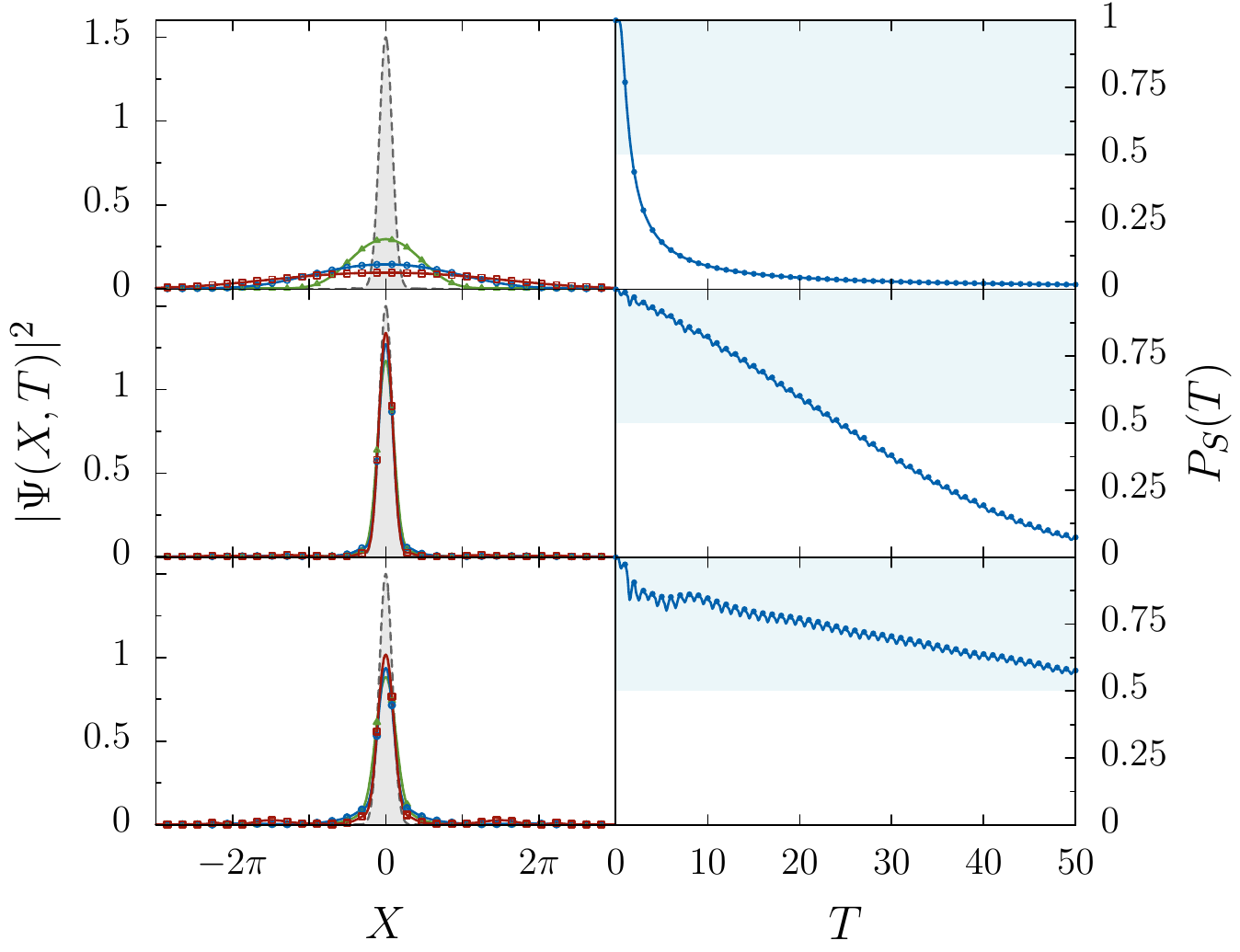}\\
\caption{Left panels show the probability density $|\Psi(X,T)|^2$ vs position $X$ at $T=0$ (single Gaussian wave packet of rms width $\tilde{\sigma}=200^{-1/4}\approx 0.27$, dashed curve delimiting gray shaded area), $T=1$ (green solid curve with triangles), $T=2$ (blue solid curve with circles), $T=3$ (red solid curve with squares) for $R_{0\ell}=0.6$ and  $Q=1$. Right panels show $P_S(T)=\int_{-\pi/2}^{+\pi/2} |\Psi(X,T)|^2 dX$. Top row shows without driving ($s=0$) and with $\g/Q=0.4$. Middle row shows with driving given by  $R_{i0,\mathrm{eff}} \approx 0.16$ at $s=200$ and $\g/Q=0.008$. 
Bottom row shows with driving corresponding to $R_{i0,\mathrm{eff}} \approx 1.09$ at $s=200$ and $\g/Q=0.4$ (left cross in Fig.~\ref{densityplotPK}).
\label{psi2andPS}}
\end{figure}

\begin{figure}
\includegraphics[width=0.87\linewidth]{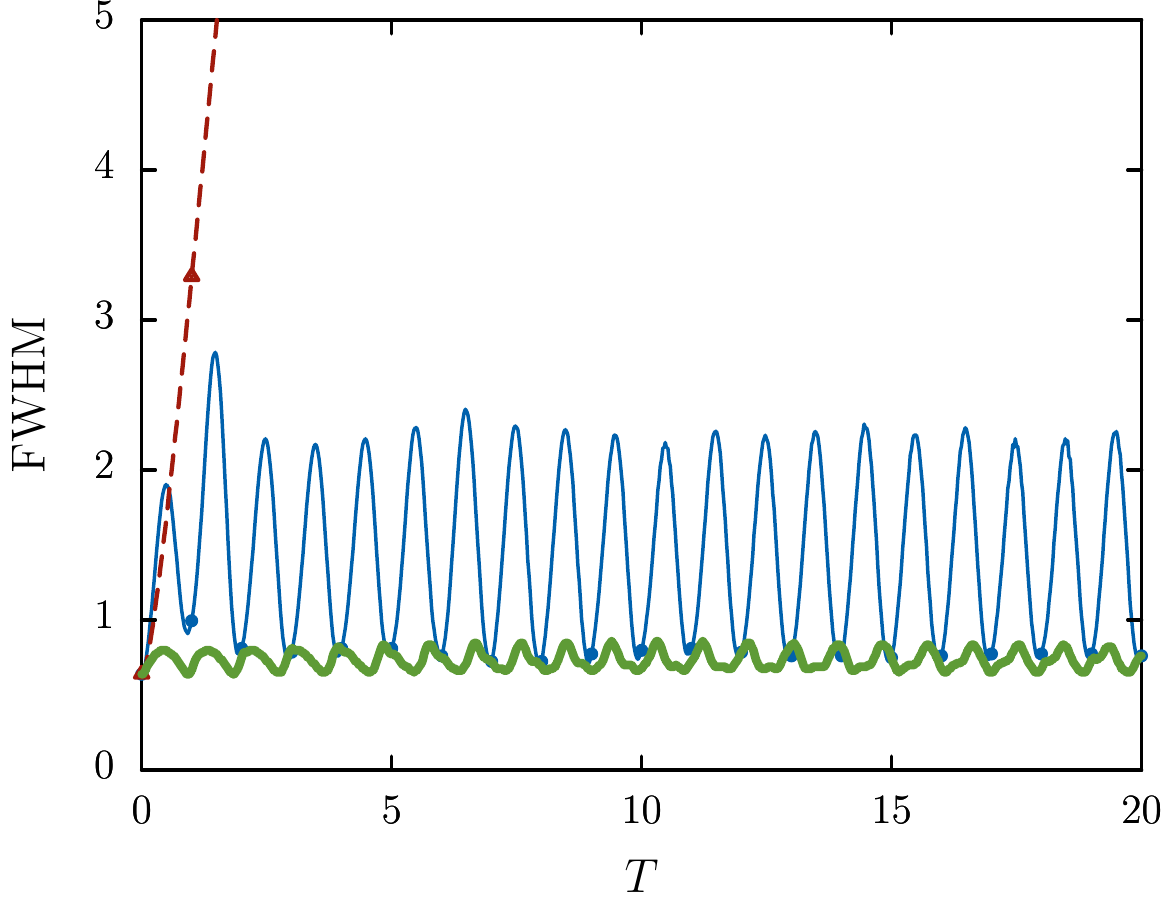}\\
\caption{Full width at half maximum (FWHM) of $|\Psi(X,T)|^2$ as a function of time $T$ for an initial Gaussian wave packet of rms width $\tilde{\sigma}=200^{-1/4}\approx 0.27$ for $R_{0\ell}=0.6$ and $\g/Q=0.4$ (with $Q=1$). Red dashed curve is without driving ; green thick solid curve is without driving but with static potential $-V_0(X)$ at $s=200$ ; blue solid curve is with driving corresponding to $R_{i0,\mathrm{eff}} \approx 1.09$ at $s=200$.
\label{FWHM}}
\end{figure}

\begin{figure}
\includegraphics[width=\linewidth]{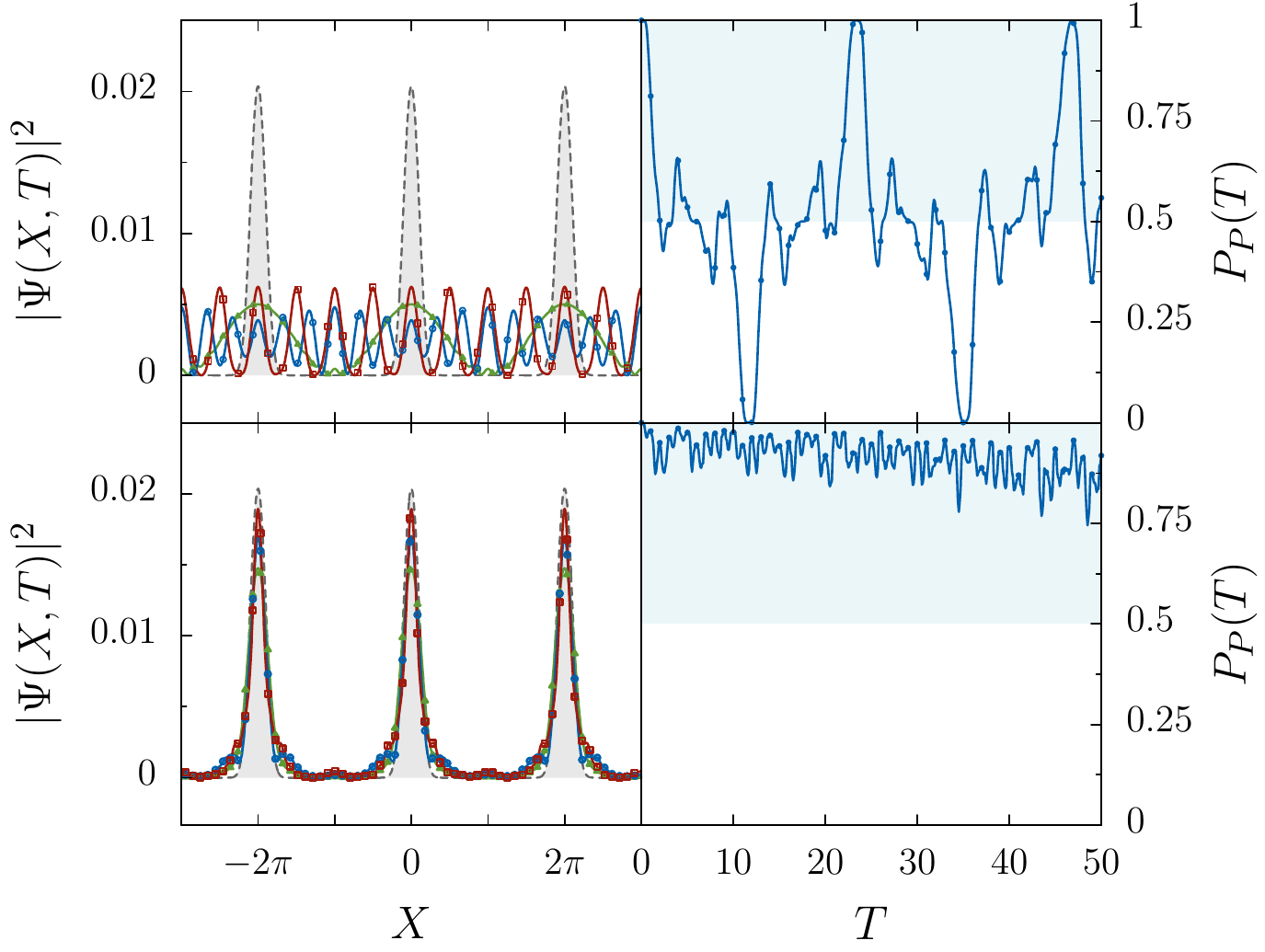}\\
\caption{Left panels show probability density $|\Psi(X,T)|^2$ vs position $X$ at $T=0$ (periodic ground-state wave function of GPE with static potential $-V_0(X)$, dashed curve delimiting the gray shaded area, here $\tilde{\sigma}\approx 0.40$), $T=1$ (green solid curve with triangles), $T=2$ (blue solid curve with circles), $T=3$ (red solid curve with squares) for $R_{0\ell}= 0.6$ and $\g/Q=0.4$ with $Q=64$. Right panels show $P_P(T)=\int_{\{X:V_0(X)>0\}} |\Psi(X,T)|^2 dX$. Top row shows without driving ($s=0$). Bottom row shows with driving given by $R_{i0,\mathrm{eff}} \approx 0.85$ and $s=200$.
\label{psi2andPP}}
\end{figure}

\section{Kapitza stabilization of quantum states}

The time evolution of a single wave packet and its full width at half maximum (FWHM) are shown in Figs.~\ref{psi2andPS} and~\ref{FWHM}, respectively.
Without the oscillating potential, the wave packet spreads over the whole lattice, leading to a monotonic drop of the probability inside the initial potential well, $P_S(T)=\int_{-\pi/2}^{+\pi/2} |\Psi(X,T)|^2 dX$, and a linear increase of the FWHM with time. 
In contrast, in the presence of the oscillating potential, the Kapitza stabilization leads to conservation of a large part of the probability in the initial well. The larger the  interaction strength $\g$, the better the stabilization.  Interestingly enough, quantum stabilization exists inside the classical stability domain (see middle panels), but also at $ R_{i0,\mathrm{eff}}\approx 1.09$, significantly above the classical stability border $ R_{i0,\mathrm{eff}}=0.21$ from Eq.~(\ref{kapitzaborder}). We attribute this quantum enhancement of Kapitza stabilization to the presence of oscillations in the width of the wave packet (see Fig.~\ref{FWHM}), which
we discuss below.

\begin{figure}
\includegraphics[width=0.85\linewidth]{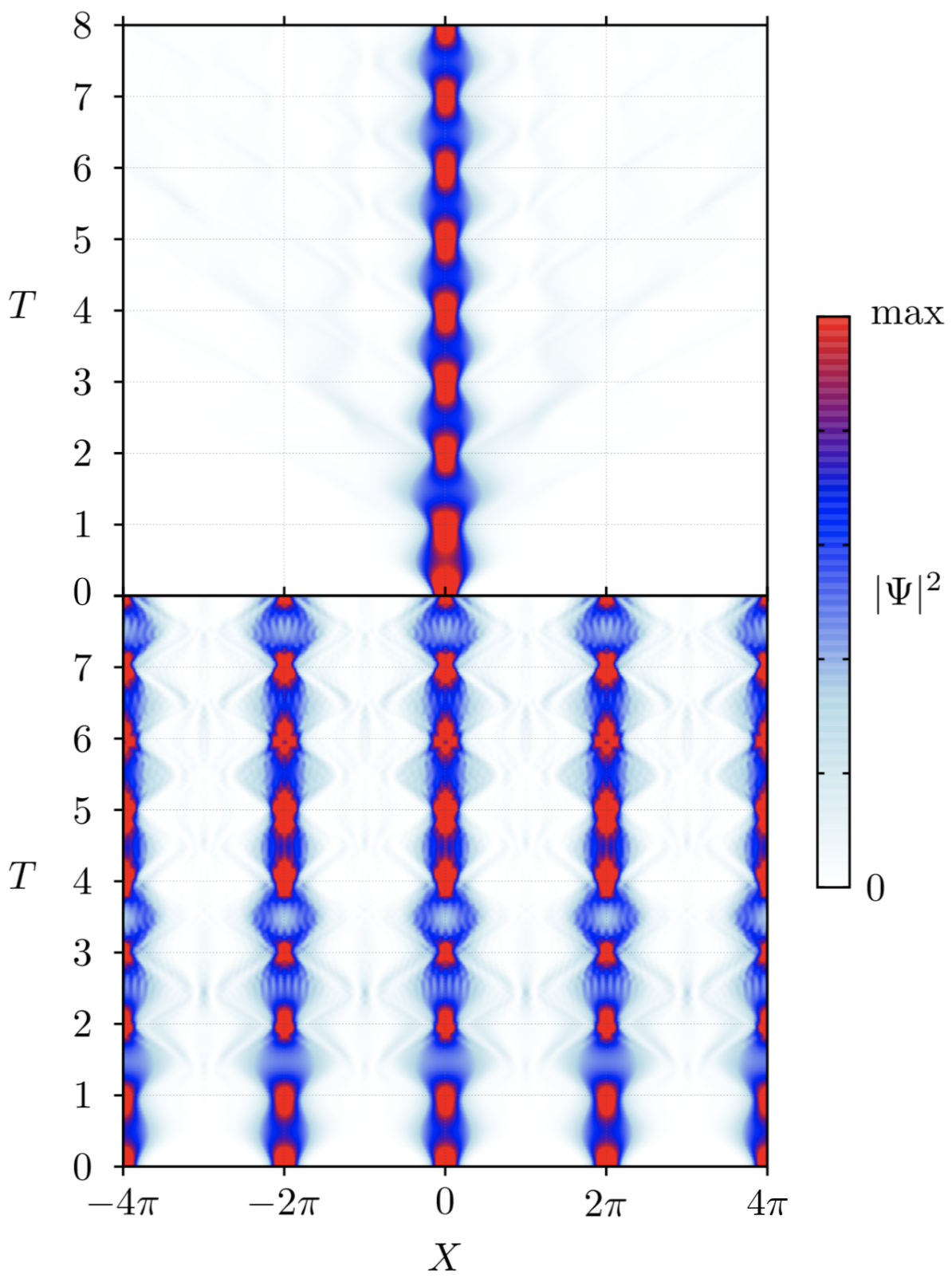}
\caption{Density plot of probability density as a function of position $X$ and time $T$ for $R_{0\ell}= 0.6$, $\g/Q=0.6$, and $s=200$. Top panel shows that the initial state is a single Gaussian wave packet of rms width $\tilde{\sigma}=200^{-1/4}\approx 0.27$ centered on $X=0$ ($Q=1$), with $R_{i0,\mathrm{eff}} \approx 1.34 $ (see right cross in Fig.~\ref{densityplotPK}). Bottom panel shows that the initial state is the periodic ground-state wave function of GPE with static potential $-V_0(X)$ ($Q=64$) and $R_{i0,\mathrm{eff}} \approx 1.03$.
\label{densityplotpsi2}}
\end{figure}

Another possible initial state is given by a chain of BEC wave packets corresponding to the ground state of GPE at each potential minimum of the lattice.
This state is obtained numerically by the standard method of imaginary time propagation of GPE. The time evolution is shown in Fig.~\ref{psi2andPP}. 
Without the oscillating potential, the periodic peak structure becomes less pronounced, decreasing with time. Thus the probability escapes from the vicinity of the unstable fixed points as measured by $P_P(T)=\int_{\{X:V_0(X)>0\}} |\Psi(X,T)|^2 dX$, which oscillates in time between $0$ and $1$. In contrast, in the presence of the oscillating potential, the probability remains in the vicinity of the unstable fixed points, even if the parameter $R_{i0,\mathrm{eff}}\approx 0.85$ is significantly beyond the classical stability border of Eq.~(\ref{kapitzaborder}). 

The origin of the quantum enhancement of the Kapitza stabilization seen in Figs.~\ref{psi2andPS} and \ref{psi2andPP} can be understood from the typical evolution of the wave function shown in Fig.~\ref{densityplotpsi2}. Indeed, the width $\tilde{\sigma}$ of the wave packet oscillates in time by a factor $f \approx 2$ (see Fig.~\ref{FWHM}), which renormalizes $\tilde{\sigma}$. Since $R_{i0,\mathrm{eff}} \propto \tilde{\sigma}^{-3/2}$, this gives a reduction factor $f^{-3/2}$ of the values of $R_{i0,\mathrm{eff}}$ in Fig.~\ref{psi2andPS} from $R_{i0,\mathrm{eff}}=1.09$ to $R_{i0,\mathrm{eff}}=0.39$, significantly closer to the theoretical classical border of Eq.~(\ref{kapitzaborder}) at $R_{i0,\mathrm{eff}}=0.21$.  Moreover, the time oscillation of $|\Psi(X,T)|^2$ creates a supplementary oscillating potential which can generate an additional Kapitza stabilization.

\begin{figure}
\includegraphics[width=0.95\linewidth]{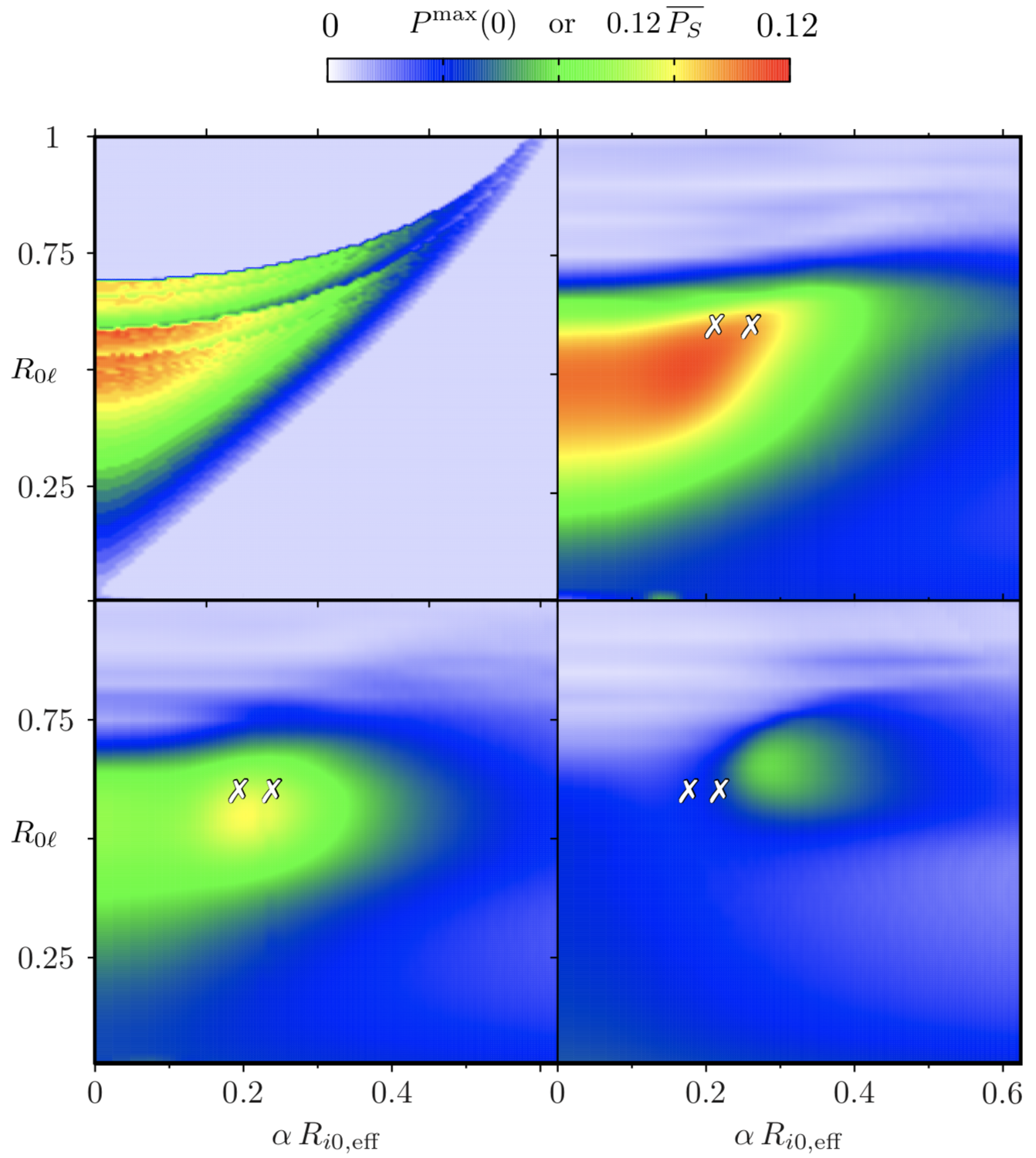}
\caption{Top left panel shows classical stability map as in Fig.~\ref{poincare_stability}. Top right and bottom panels show density plot of $\overline{P_S}=\tfrac{1}{10}\int_{3T_{\mathrm{sp}}}^{3T_{\mathrm{sp}}+10}P_S(T)dT$, with $P_S(T)=\int_{-\pi/2}^{+\pi/2} |\Psi(X,T)|^2dX$ and where $T_{\mathrm{sp}}$ is a characteristic spreading time of the initial wave packet in the absence of driving ($s=0$) determined by $P_S^{s=0}(T_{\mathrm{sp}})=0.75$, as a function of the frequency ratio $R_{0\ell}$  and the effective frequency ratio $R_{i0,\mathrm{eff}}$ of Eq.~(\ref{effRi0}) (multiplied by a scaling coefficient $\alpha=0.2$) for $s=200$ (top right), $s=100$ (bottom left) and $s=50$ (bottom right). Initial state is a single Gaussian wave packet of rms width $\tilde{\sigma}=1/s^{1/4}$ centered around $X=0$ ($Q=1$). The two crosses correspond to parameters $\g/Q=0.4$ (left cross) and $\g/Q=0.6$ (right cross) for $R_{0\ell}=0.6$. \label{densityplotPK}}
\end{figure}

The region of quantum Kapitza stabilization in this GPE system is shown in Fig.~\ref{densityplotPK}, which displays the time-averaged probability to stay in the vicinity of the unstable fixed point $X=0$ as a function of $R_{i0,\mathrm{eff}}$ and $R_{0\ell}$ together with the classical stability diagram of Fig.~\ref{poincare_stability}. In the regime of small effective Planck's constant corresponding to large $s$ values, a large stability region is visible, with a shape similar to the classical stability domain. In Fig.~\ref{densityplotPK}, the values of $R_{i0,\mathrm{eff}}$ are rescaled
by a multiplicative factor $\alpha=0.2$ corresponding to the fact that quantum stabilization exists at $R_{i0,\mathrm{eff}}$ values significantly larger than in the classical case, given by Eq.~(\ref{kapitzaborder}).  We attribute the presence of this factor to the quantum fluctuations as discussed above. For decreasing values of $s$, the quantum stability region becomes less pronounced. We explain this by the fact that the effective $\hbar$ becomes comparable to the phase-space area of the classical Kapitza stability island (see Fig.~\ref{poincare_stability}, top left panel). In this case, quantum tunneling from the island becomes important and leads to the destruction of the Kapitza phenomenon.

We note that in the absence of interaction, quantum tunneling will always induce a decay of probability inside the stability island induced by Kapitza stabilization. However, this tunneling can be affected in a nontrivial way by the nonlinearities. Indeed, nonlinearities can produce solitonic solutions which remain stable for all times. Therefore, a rigorous answer to this interesting question requires a careful mathematical analysis. All time stable solutions may appear when the stability region has very large depth and sufficiently large width. 

\section{Proposed experimental realization}

The experimental implementation can be carried out by loading adiabatically a BEC into a deep static horizontal 1D optical lattice ($s\sim 50$) realized with far off-resonant lasers.  As a result, we obtain a chain of small BEC at the bottom of the potential wells. To place them at the top of the potential hills of the lattice, we have to shift suddenly by half the spatial period the optical lattice as in Ref.~\cite{For16}. The amplitude of the lattice shall be subsequently modulated to ensure the Kapitza stabilization. In practice, the control of the lattice parameters (phase, amplitude) can be performed by using phase-locked synthesizers that imprint their signals on light through  acousto-optic modulators (AOMs) placed on each lattice beam before they interfere to produce the lattice. The range of interaction strengths that we propose is readily achievable with a standard rubidium-87 BEC placed in an optical lattice made of two counterpropagating lasers at 1064 nm. With $\omega_{\perp}\approx 2\pi\times 200$~Hz and a lattice spacing $d\approx 532$~nm, $\g\approx 0.003 N/\sqrt{s}$, we have $\g\approx 21$ for $N=10^{5}$ and a depth $s=200$, and $\g/Q\simeq 0.55$ for a BEC of typical size 20 $\mu$m. Interestingly, the enhancement of interactions through Feshbach resonances is not necessary to observe the dynamical stabilization phenomenon. We note that Kapitza stabilization of cold atoms in optical lattices starts to attract the interest of experimental groups \cite{weld}.

\section{Conclusion}  We have shown that the Kapitza phenomenon can stabilize a BEC with repulsive interaction by means of an oscillating force with zero average. This represents an application of the Kapitza effect in the context of nonlinear quantum fields. Our theoretical proposal can be experimentally realized with current cold atom technology. Besides its fundamental interest, it should provide new tools for the long-time manipulation of BECs.

\begin{acknowledgments}
This work was supported in part by the Programme Investissements
d'Avenir ANR-11-IDEX-0002-02, reference ANR-10-LABX-0037-NEXT (projects THETRACOM and TRAFIC). 
Computational resources were provided by the Consortium des Equipements de Calcul Intensif (CECI), 
funded by the Fonds de la Recherche Scientifique de Belgique (F.R.S.-FNRS) 
under Grant No.~2.5020.11 and by CAlcul en MIdi-Pyr\'en\'ees (CALMIP).
\end{acknowledgments}


\end {document}